\documentclass[twocolumn,twocolappendix,trackchanges]{aastex631}
\usepackage{amsmath}
\hypersetup{
	colorlinks	= true,
	linkcolor	= red,
	urlcolor	= cyan,
	citecolor	= blue
}

\usepackage{amssymb}
\usepackage{xcolor}
\definecolor{mygreen}{rgb}{0.1, 0.7, 0.2}
\definecolor{myorange}{rgb}{1,0.5,0}
\definecolor{myred}{rgb}{1,0,0}

\usepackage{lipsum}
\usepackage{multirow}

\graphicspath{{./Figures/}}

\newcommand{\exorelr}{\mbox{\textsc{ExoReL$^\Re$}}}

\received{November 4$^{th}$, 2024}
\accepted{December 28$^{th}$,  2024}

\submitjournal{AJ}

\shortauthors{Damiano et al.}

\usepackage{fancyhdr}
\begin{document}

\title{Effects of planetary mass uncertainties on the interpretation of the reflectance spectra of Earth-like exoplanets}

\correspondingauthor{Mario Damiano}
\email{mario.damiano@jpl.nasa.gov}

\author[0000-0002-1830-8260]{Mario Damiano}
\affiliation{Jet Propulsion Laboratory, California Institute of Technology, Pasadena, CA 91109, USA}

\author[0009-0000-4845-3613]{Zachary Burr}
\affiliation{Jet Propulsion Laboratory, California Institute of Technology, Pasadena, CA 91109, USA}

\author[0000-0003-2215-8485]{Renyu Hu}
\affiliation{Jet Propulsion Laboratory, California Institute of Technology, Pasadena, CA 91109, USA}
\affiliation{Division of Geological and Planetary Sciences, California Institute of Technology, Pasadena, CA 91125, USA}


\author[0000-0002-0040-6815]{Jennifer Burt}
\affiliation{Jet Propulsion Laboratory, California Institute of Technology, Pasadena, CA 91109, USA}

\author[0000-0003-3759-9080]{Tiffany Kataria}
\affiliation{Jet Propulsion Laboratory, California Institute of Technology, Pasadena, CA 91109, USA}




\begin{abstract}

Atmospheric characterization of Earth-like exoplanets through reflected light spectroscopy is a key goal for upcoming direct imaging missions. A critical challenge in this endeavor is the accurate determination of planetary mass, which may influence the measurement of atmospheric compositions and the identification of potential biosignatures. In this study, we used the Bayesian retrieval framework \exorelr\ to investigate the impact of planetary mass uncertainties on the atmospheric characterization of terrestrial exoplanets observed in reflected light. Our results indicate that precise prior knowledge of the planetary mass can be crucial for accurate atmospheric retrievals if clouds are present in the atmosphere. When the planetary mass is known within 10\% uncertainty, our retrievals successfully identified the background atmospheric gas and accurately constrained atmospheric parameters together with clouds. However, with less constrained or unknown planetary mass, we observed significant biases, particularly in the misidentification of the dominant atmospheric gas. For instance, the dominant gas was incorrectly identified as oxygen for a modern-Earth-like planet or carbon dioxide for an Archean-Earth-like planet, potentially leading to erroneous assessments of planetary habitability and biosignatures. These biases arise because, the uncertainties in planetary mass affect the determination of surface gravity and atmospheric scale height, leading the retrieval algorithm to compensate by adjusting the atmospheric composition. Our findings emphasize the importance of achieving precise mass measurements—ideally within 10\% uncertainty—through methods such as extreme precision radial velocity or astrometry, especially for future missions like the Habitable Worlds Observatory.


\end{abstract}




\section{Introduction} \label{sec:intro}
The search for potentially habitable worlds has become a paramount objective in space research. Identifying planets that could host life is not only a fascinating scientific endeavor but also a crucial step towards understanding our place in the universe. In line with this goal, the Astro2020 decadal survey has prioritized the development of a 6-meter class telescope designed to image habitable-zone terrestrial exoplanets in reflected light, highlighting its significance for future missions \citep{astro2020national}. This telescope, now named ``Habitable Worlds Observatory,'' would enable the detection and detailed characterization of potentially Earth-like planets with unprecedented precision, providing insights into their atmospheric composition, surface conditions, and potential habitability.


Accurate knowledge of the planetary mass is shown to be crucial for determining the composition of exoplanetary atmospheres using the transit method, the currently dominating method to characterize exoplanets \citep{batalha2019mass}. The mass of a planet influences its gravitational pull, atmospheric pressure, and the distribution of gases, all of which affect the observed spectra. Similarly, the knowledge of the planetary mass is presumed to be important for the proper characterization of planets via reflected light \citep{eprv2021}, but the specific dependency of the measurement ability on the planetary mass uncertainties remains unclear. 

To address this issue, we explore the impact of planetary mass uncertainties on the atmospheric retrieval of terrestrial exoplanets observed in reflected light. We use \exorelr\ \citep{damiano2020exorel,damiano2022small,damiano2023uv} to interpret the spectra of two different atmospheres, (i.e., Modern and Archean Earth analogs) and we assess the effect of different planetary mass prior knowledge on the retrieved posterior probability distribution functions.

In the context of reflection spectroscopy, \cite{salvador2024mass} discussed the impact of orbit and mass uncertainties. \cite{salvador2024mass} investigated the impact of orbital and mass constraints, derived from precursor radial velocity surveys or astrometry, on the retrievals of small terrestrial planetary properties. They suggested that the prior information on orbital parameters significantly tightens constraints on planetary radius, and additional mass knowledge does not notably enhance the retrieval of atmospheric or bulk properties. Conversely, our study focuses on the effects of planetary mass and radius uncertainties on the retrievable atmospheric information. \cite{salvador2024mass} highlighted that higher signal-to-noise ratio (SNR) could be more beneficial than additional prior mass information for atmospheric characterization. While we do not explore a range of S/N scenarios, our study on the interplay between the mass uncertainty and spectral information complements the work presented in \cite{salvador2024mass}.

\exorelr\ has already proven to be a valuable tool for characterizing terrestrial exoplanets in reflected light \citep{damiano2022small}. This study builds on that foundation by introducing several enhancements to the retrieval software and addressing key scientific questions. Our retrieval framework incorporates advanced statistical methods to derive atmospheric parameters from spectral data, preserving the causal relationships between observed spectral features and underlying atmospheric properties. By applying \exorelr, we aim to quantify the impact of mass and radius uncertainties on the retrieval accuracy and identify the most valuable precursor observational parameters for future missions.

This paper is structured as follows: Section \ref{sec:method} discusses the setup of the retrievals, key upgrades to \exorelr, and the various atmospheric scenarios explored in this study. In Section \ref{sec:results}, we present the retrieval results, highlighting the influence of mass and radius uncertainties on the derived atmospheric properties. Section \ref{sec:discuss} articulates the implications of these findings for the development of future missions. Finally, Section \ref{sec:conclusion} summarizes the key points of this work, emphasizing the importance of accurate mass and radius measurements for the successful characterization of terrestrial exoplanets in reflected light.
\section{Methods}\label{sec:method}

We used \exorelr\ to explore the relationship between mass-radius uncertainties and posterior constraints from retrievals of reflected light spectra. \exorelr\ is a Bayesian inverse retrieval framework for spectroscopically direct imaged exoplanets \citep{damiano2020exorel,damiano2022small,damiano2023uv}. A radiative transfer model is used to generate spectra which are compared with the observed spectra. The free parameters are systematically varied to determine what combination will produce a spectra closest to the observed spectra. In this way, key parameters such as gas abundances and the size of the planet can be determined from its spectrum.

\subsection{Upgrades to \exorelr} \label{subsec:method_upgrades}
Compared to our previous works, we introduced a series of upgrades to the \exorelr\ software to carry out this study. 

\subsubsection{Cloud Fraction}\label{subsec:cld_frac}
The cloud fraction has been introduced as a parameter in our forward model. Although cloud density is self consistently calculated within \exorelr\ so that the transparency of the cloud can be modeled, the cloud coverage of the integrated planetary disk was assumed to be 100\% \citep{damiano2020exorel}. Previously, the cloud transparency/density could not reproduce the behavior patchy clouds as the cloud formulation carries spectral features from water vapor condensation. Now, we calculate two models: one with the original cloud formulation and the other that is free from clouds. We take the average of the two models, weighted by the cloud fraction, as the modeled reflected spectrum. This change effectively moves \exorelr\ from a 1D model to 1.5D model as now two atmospheric columns are taken into consideration to calculate the model reflected light. In this manuscript, we fixed the cloud fraction to 25\%. The effect of having cloud fraction as a free parameter will be explored in a companion study (Burr et al., in prep.).

\subsubsection{Rayleigh Scattering}\label{subsec:ray}
Another upgrade we introduced is the ability to calculate the Rayleigh scattering according to the atmospheric composition. The contribution to the Rayleigh scattering of each of the gases is taken into account and the total Rayleigh scattering is the weighted average of the Rayleigh scattering due to each of the individual gases based on the volume mixing ratios of the gases. The Rayleigh scattering is important to constrain the dominant gas of the atmosphere; especially if the gas does not show significant spectral features in the considered wavelength range (NUV, VIS, and NIR), e.g., $N_2$ and $H_2$. Additionally, the Rayleigh scattering is sensitive to the amount of clouds as these tend to mute the spectral feature of Rayleigh scattering. When we include a cloud-free column in the forward model, a Rayleigh slope in the spectrum will typically result in a strong reflection at shorter wavelength and thus the need for ``retrieving'' Rayleigh scattering.

\subsubsection{Adaptive vertical grid of model atmosphere}\label{subsec:grid}
In the context of radiative transfer calculation, the layering of the vertical profile plays a crucial role. The more layers the atmosphere is divided into, the more precise the calculation of the energy transfer will be, resulting in a more precise synthesized spectrum. However, the bigger the number of layers, the more computationally intensive the calculation is. In the context of retrieval calculations a balance between these two effects should be employed to reach good precision in the calculations while being sufficiently fast.
Generally, the layers are equally spaced in the planetary scale height and when translated into physical height more layers are located in the upper part of the atmosphere. While this is a good practice to capture the absorption features of the atmospheric gases, it might create a situation in which the clouds and the atmosphere beneath are poorly sampled which will create distortion in the calculated albedo spectrum. In literature, e.g., \citet{batalha2019picaso}, this issue has been solved by linking the number of layers in which the cloud is defined with the optical thickness of the cloud ($\tau_c$). In \exorelr, $\tau_c$ is not a free parameter, and so we decided to approach the problem from a different angle. We defined a grid of layers with the same number of layers above, within, and below the cloud layer regardless of the vertical position of the clouds. In this way, we always ensure that there are enough layers to fully describe radiative transfer within the cloud layer. We refer to this new scheme in this manuscript as the ``\textit{adaptive grid}.'' For the rest of this manuscript, we used 80 layers in this adaptive grid scheme. This new scheme allows us to reduce the computation time as each of the spectra in the retrieval do not require a large number of layers (200$+$). As an example, with this new scheme we observed that the adaptive grid with 100 layers produce the same spectrum as a standard grid with 500 layers.

\subsubsection{Partial Pressure Sampling}\label{subsec:pp}
In literature, when retrieving the concentration of molecular species, the volume mixing ratio (VMR) is generally the most direct expression that can be used. This scheme works well when giant planets are considered as $H_2$ is implicitly adopted as the dominant gas of the atmosphere. When moving towards small planets a wider range of bulk atmospheric compositions needs to be taken into consideration and the dominant gas is not trivial \citep[e.g.,][]{hu2014small}. In this regime, the centered-log-ratio (CLR) can be used to leverage its advantage on compositional analysis \citep{beneke2012retrieval, damiano2021prior}. These two methods have worked well for planets observed via transmission and/or emission spectroscopy. In reflection spectroscopy, we do not define the radius of the planet at a specific pressure (as is done in transmission spectroscopy, for example) because we are sensitive to the surface pressure which is linked with the planetary radius. In this work, we did not fit the surface pressure and instead we introduced the partial pressure of the gases as free parameters \citep{salvador2024mass}. The sum of the gas partial pressures is the surface pressure by definition (see Equations \ref{eq:pp1} and \ref{eq:pp2}):

\begin{equation}
    P_0 = \sum_{n} \delta P_{gas_n}
    \label{eq:pp1}
\end{equation}

\begin{equation}
    \delta P_{gas_n} = P_0 \times VMR_{gas_n}
    \label{eq:pp2}
\end{equation}

\noindent where $\delta P_{gas_n}$ is the partial pressure of the $nth$ gas, $VMR_{gas_n}$ is the $nth$ gas volume mixing ratio, and $P_0$ is the surface pressure.

Using the partial pressure to define the atmospheric composition gives the advantage of removing the need for the concept of a \textit{filler gas} as all the gas partial pressures are free and independent parameters, including those that are not spectroscopically active in the considered wavelength range. Another advantage is the ease of switching from the partial pressure space to the VMR space as the transformation is linear. When CLR is used, the transformation to VMR is not linear, and non-uniform priors may be required \citep{damiano2021prior}.

\subsubsection{Planetary Mass as Free Parameter}\label{subsec:mass_2Dprior}
\begin{figure}
    \centering
    \includegraphics[width=\linewidth]{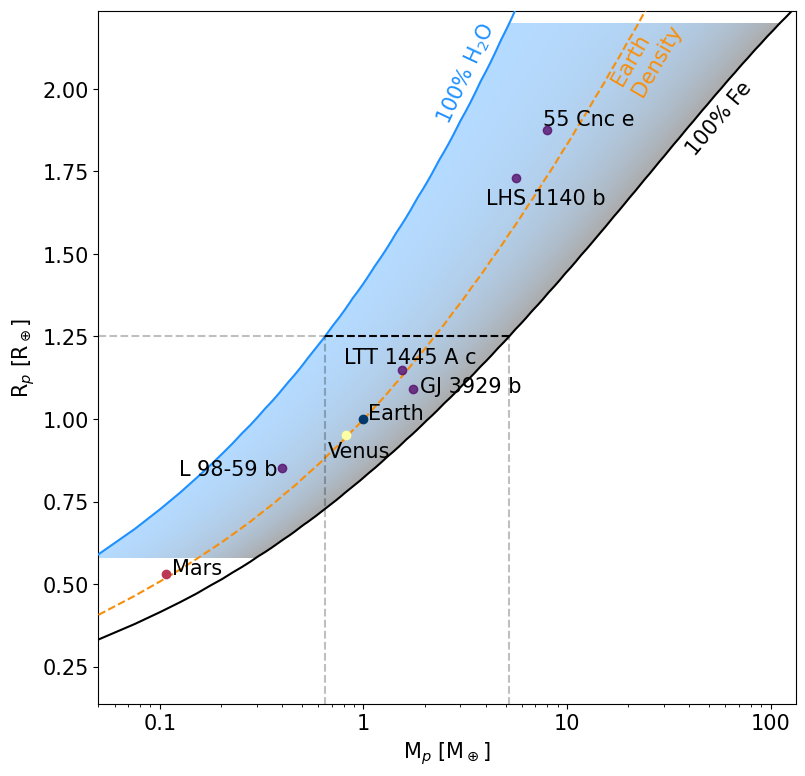}
    \caption{Mass radius relations used for the 2D prior \citep{zeng2016mass}. The shaded region shows the full search space of the retrieval when using this prior function. The radius is chosen from a uniform distribution between 0.58 and 2.2 $R_{\oplus}$, then the mass is chosen from a uniform distribution between the corresponding mass for a 100\% water ice planet and a 100\% iron planet with this radius. For example for a radius of 1.25 $R_{\oplus}$, the mass can vary from 0.65 to 5.2 $M_{\oplus}$.}
    \label{fig:MR-relations}
\end{figure}

Given the focus on this parameter in this study, we defined and explored different prior probability functions. 
A novel type of prior function, the ``\textit{2D prior}", is introduced which leverages the correlation between planetary mass and radius. This conditional mass-radius prior relation was adopted from \cite{zeng2016mass} models (see Figure \ref{fig:MR-relations}). The planetary radius has a uniform prior, and for a given radius, the mass is drawn from a uniform distribution between the corresponding mass for a 100\% iron planet to the corresponding mass for a 100\% water ice planet. These two cases represent the theoretical upper and lower bounds on density for a terrestrial planet. This 2D prior limits the search space to only physically realistic scenarios, enabling better fitting of the planetary mass than the independent uniform prior. By using this prior function, we are introducing information into our retrieval process that is physically motivated. When using the 2D prior, the uniform distribution for radius is decreased from [0.5, 10] $R_{\oplus}$ to [0.58, 2.2] $R_{\oplus}$, as this is the range that the models of \citet{zeng2016mass} cover.

Additionally, we also consider the simplest prior for the planetary mass: a uniform flat prior between two fixed values ([0.01, 20] $M_{\oplus}$), as well as a Gaussian prior to simulate a situation where the mass uncertainty is provided (e.g., with radial velocity measurements). In this study, we considered two Gaussian priors that corresponds to a 3.3$\sigma$ or 10$\sigma$ detection of the planetary mass. 

\subsection{Retrieval Setup}\label{subsec:method_setup}
In this study, the partial pressure of H$_2$O, CH$_4$, CO$_2$, O$_2$, O$_3$, and N$_2$ are considered as free parameters, and we transform the partial pressures into VMR in the posterior distribution graphs shown in this paper. We consider water clouds as the only condensates. The clouds are modeled as described in \citet{damiano2020exorel}, and are fit with the cloud top pressure ($P_{top,H_2O}$), cloud depth ($D_{cld,H_2O}$), and condensation ratio ($CR_{H_2O}$), as well as the VMR of water below the clouds. We also included a variation in which the cloud particle size is additionally considered as a free parameter. In \exorelr, the particle size is typically estimated based on microphysics \citep{hu2019information}. By including a free particle size, we assume instead a constant average value across the atmosphere. The planetary mass and the planetary radius are included as free parameters as described in Section \ref{subsec:mass_2Dprior}. 

Table~\ref{tab:retrieval_setup} lists the free parameters, the prior space used, and the range in which the parameters are probed. \exorelr\ uses \texttt{MultiNest} \citep{feroz2009multinest} to sample the Bayesian evidence, estimate the parameters, and calculate the posterior distribution functions. \texttt{MultiNest} is used through its \texttt{Python} implementation \texttt{pymultinest} \citep{buchner2014multinest}. For all the retrieval analyses presented here, we used 2000 live points and 0.5 as the Bayesian evidence tolerance.

\begin{deluxetable}{lcl}[!h]
    \tablecaption{Model parameters and prior probability distributions used in the atmospheric retrievals. $\mathcal{U}(a,b)$ is the uniform distribution between values $a$ and $b$, $\mathcal{LU}(a,b)$ is the log-uniform (Jeffreys) distribution between values $a$ and $b$, and $\mathcal{N}(\mu,\sigma^2)$ is the normal distribution with mean $\mu$ and variance $\sigma^2$. NOTE - $(^1)$ \cite{zeng2016mass}. \label{tab:retrieval_setup}}
    \tablehead{\textbf{Parameter} & \textbf{Symbol} & \textbf{Prior}}
    \startdata
        Cloud top pressure [Pa] & $P_{top,H_2O}$ & $\mathcal{LU}$(2.0, 7.0) \\
        Cloud depth [Pa] & $D_{cld,H_2O}$ & $\mathcal{LU}$(2.0, 7.0) \\
        Condensation ratio & $CR_{H_2O}$ & $\mathcal{LU}$(-7.0, 0.0)\\
        Partial pressure H$_2$O [Pa] & PP(H$_2$O) & $\mathcal{LU}$(-7.0, 7.0) \\
        Partial pressure CH$_4$ [Pa] & PP(CH$_4$) & $\mathcal{LU}$(-7.0, 7.0) \\
        Partial pressure CO$_2$ [Pa] & PP(CO$_2$) & $\mathcal{LU}$(-7.0, 7.0) \\
        Partial pressure O$_2$ [Pa] & PP(O$_2$) & $\mathcal{LU}$(-7.0, 7.0) \\
        Partial pressure O$_3$ [Pa] & PP(O$_3$) & $\mathcal{LU}$(-7.0, 7.0) \\
        Partial pressure N$_2$ [Pa] & PP(N$_2$) & $\mathcal{LU}$(-7.0, 7.0) \\
        Planetary mass [M$_{\oplus}$] & M$_p$ & $\mathcal{U}$(0.01, 20.0) \\
        & & $\mathcal{N}$(1.0, $\sigma^2_{M_p}$) \\
        & & 2D prior$(^1)$\\
        Planetary radius [R$_{\oplus}$] & R$_p$ & $\mathcal{U}$(0.5, 10.0)\\
        \multicolumn{2}{c}{\textit{(when using the 2D prior for mass)}} & $\mathcal{U}$(0.58, 2.2)\\
    \enddata
\end{deluxetable}

\subsection{Atmospheric and Observational Scenarios}\label{subsec:method_scenarios}
To study the effect of different mass uncertainties on the interpretation of the atmospheric characterization, we used three different terrestrial planet scenarios:

\begin{enumerate}
    \item A cloud-free modern Earth-like atmosphere: $N_2$ dominated ($\sim79$\%) with $O_2$ ($\sim20$\%) and trace amounts of H$_2$O, O$_3$, CH$_4$, and CO$_2$ (see \autoref{tab:modern_cld});
    \item A modern Earth-like atmosphere: N$_2$ dominated ($\sim79$\%) with O$_2$ ($\sim20$\%) and trace amounts of H$_2$O, O$_3$, CH$_4$, and CO$_2$, and a cloud deck at $\sim$0.6 bar with a cloud fraction of 25\% (see \autoref{tab:modern});
    \item An Archean Earth-like atmosphere: The Earth's atmosphere as it was approximately 4--2.5 billion years ago. Similar to the modern Earth, but with less O$_2$ and considerably more CH$_4$ and CO$_2$ (see \autoref{tab:archean}).
\end{enumerate}

For all the scenarios, 1 M$_{\oplus}$, 1 R$_{\oplus}$, 1 AU, and 60$deg$ phase angle were adopted. We considered five retrieval scenarios: a case in which the mass is fully known and the radius is a free parameter along with other atmospheric parameters, and four scenarios in which the planetary mass is a free parameter and varying different prior functions as listed in Table \ref{tab:retrieval_setup}. Additionally, for the modern Earth-like and Archean Earth-like, we ran an additional case in which the particle size is a free parameter instead of being self-consistently calculated.
The spectra are generated in the near ultra-violet (NUV) from 0.25-0.4 $\mu$m at R=7, the visible (VIS) from 0.45-1.0 $\mu$m at R=140, and ``NIR'' from 1.0-1.8 $\mu$m at R=70. This is consistent with previous mission concept studies \citep{roberge2018irouv,gaudi2020habex}.

A new noise model was developed for this work, which is explained in more detail in \autoref{sec:noise}. The model includes wavelength-dependent behavior of noises in high-contrast imaging. It represents a simple limiting case with very good starlight suppression and low detector noise, in which photon noise from the astrophysical scene dominates. We follow the analytic prescription outlined in \citet{robinson2016characterizing}. We assume exozodi dominate over local zodi, and neglect local zodi. Dark current is also ignored, assuming that a future exoplanet imaging mission will have good enough detectors that the dark current will be lower than photon noise at all wavelengths, as is already the case for the Roman CGI in broadband imaging \citep{morrissey2023photon}. The speckle residuals are likewise neglected. This leaves the shot noise from the planet and the exozodi as the only remaining noise sources. The model is scaled to an SNR of 20 at 0.75 $\mu$m. This noise model defines the errorbars of the synthesized spectra, and corresponding Gaussian noise is added to simulate a realistic observation. 
\section{Results}\label{sec:results}
\begin{figure*}[ht]
    \centering
    \includegraphics[scale=0.475]{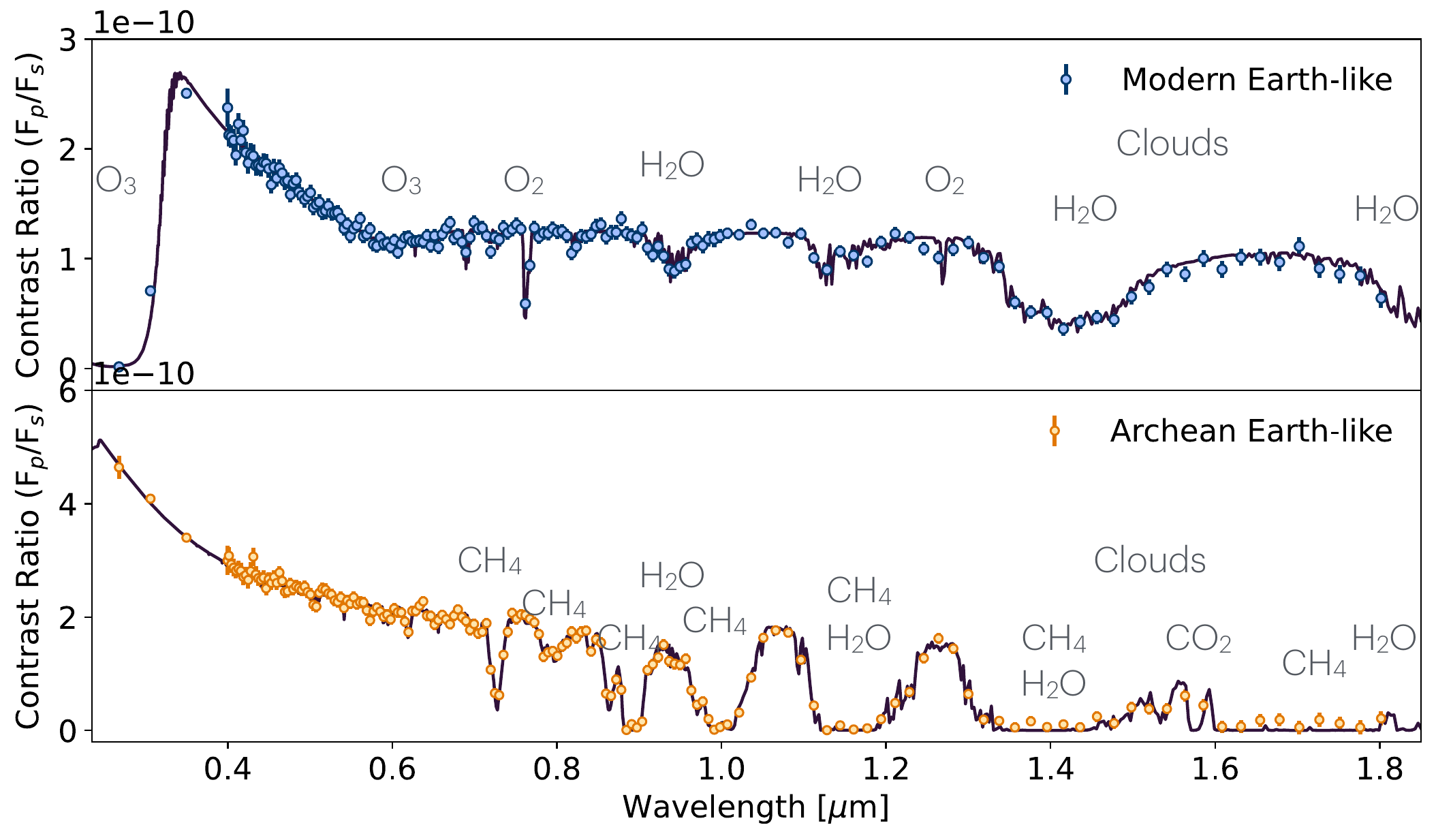}
    \caption{Simulated spectra and data models for the Modern and Archean Earth-like scenarios. The datapoints are obtained by binning the synthesized model and by applying Gaussian noise and errobars as defined in section \ref{sec:noise}. \textbf{Top panel} Modern Earth-like. \textbf{Bottom panel} Archean Earth-like.}
    \label{fig:spec}
\end{figure*}

After simulating the reflected light spectra and generating the synthetic data (Figure \ref{fig:spec}), we proceeded to run \exorelr\ for each scenarios previously described in Section \ref{sec:method}. Table \ref{tab:modern_cld}, \ref{tab:modern} and \ref{tab:archean} report the numerical values of the retrieval scenarios considered in this work.

\subsection{Cloud-free Modern Earth-like}\label{sec:mdrn_earth_nocld}

\begin{figure*}
    \centering
    \includegraphics[width=\textwidth]{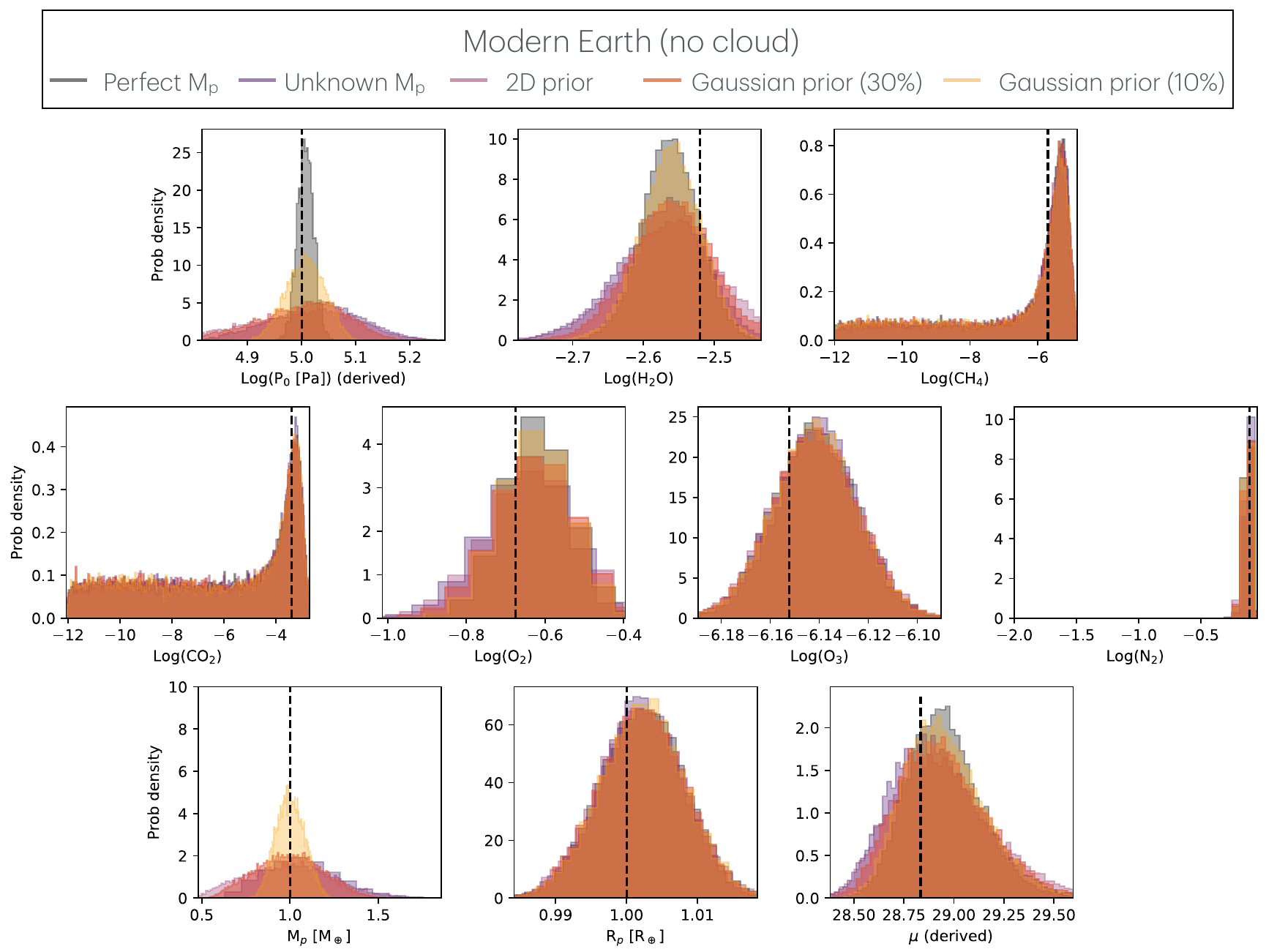}
    \caption{Posterior distribution functions of the Bayesian analysis on the cloud free Modern Earth-like reflected spectrum of all scenarios defined in \autoref{subsec:mass_2Dprior}}
    \label{fig:1D_post_modern_nocld}
\end{figure*}

Without clouds as a complicating factor, \exorelr\ is able to correctly determine all the free variables (i.e. the truth value is included within the posterior distribution functions) in all scenarios (see Figure \ref{fig:1D_post_modern_nocld}). The background gas is correctly determined to be N$_2$, with O$_2$ being the second most abundant. The radius is very tightly constrained (to within 1\%) in all scenarios. 

\begin{figure}
    \centering
    \includegraphics[width=\linewidth]{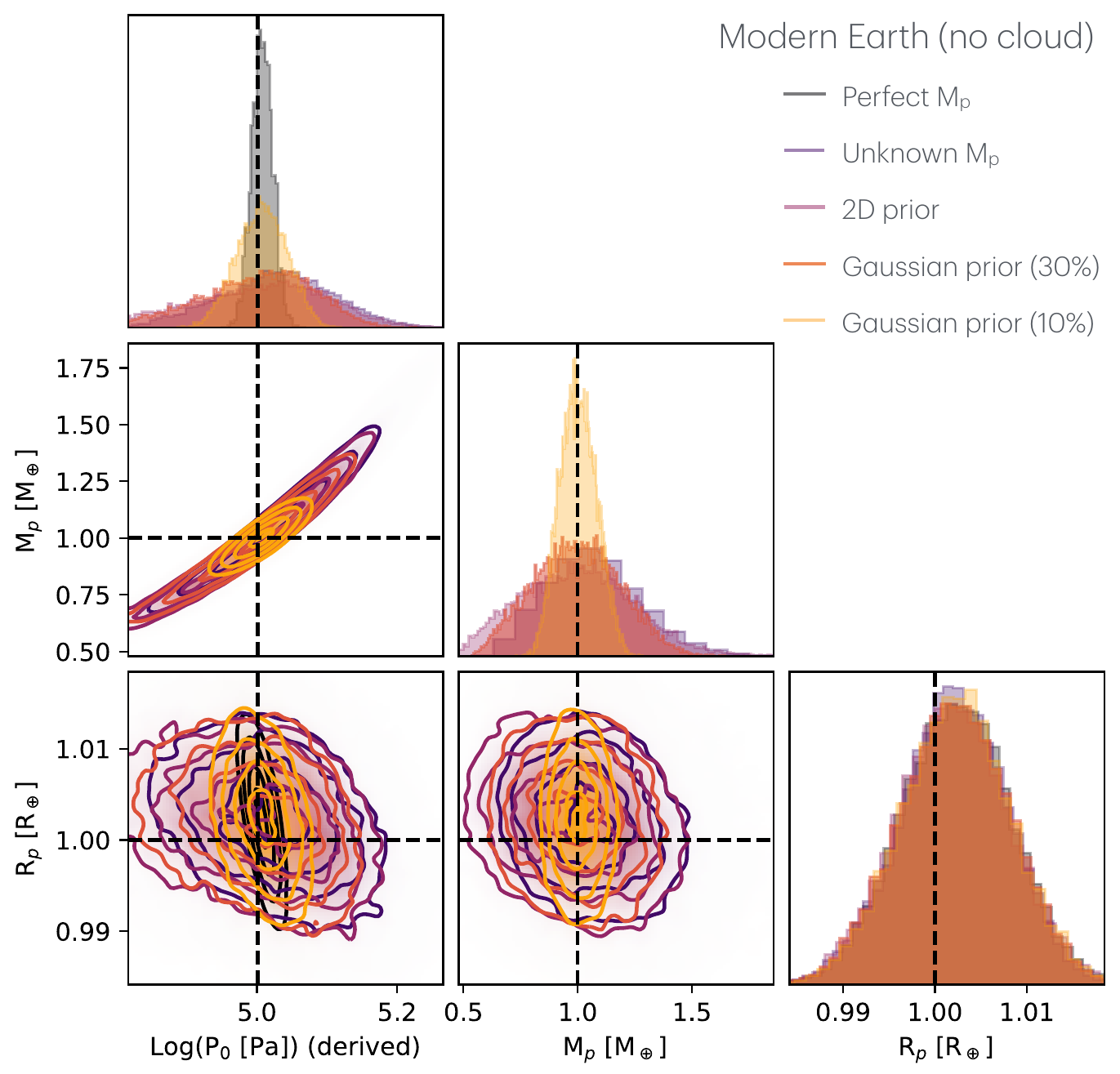}
    \caption{Posterior distribution functions for planetary mass, radius, and surface pressure for the cloud free Modern Earth-like scenario. The surface pressure is a proxy for the atmospheric gases collectively as the sum of their partial pressure is equal to the surface pressure.}
    \label{fig:mdrn_post_nocld}
\end{figure}

Zooming in on just the planetary mass, radius, and surface pressure (Figure \ref{fig:mdrn_post_nocld}), we do not find correlations between planetary mass and radius. However, there is a strong positive correlation between mass and surface pressure, indicating that spectrum is approximately sensitive to the column abundances ($\sim P/g$). For the unknown prior, the 2D prior, and the 30\% Gaussian prior, the mass is constrained to $1\pm0.2\ M_{\oplus}$. This shows that some knowledge of the mass can be derived from the spectrum itself. The 10\% Gaussian prior has tighter constraints on the mass than what can be derived from the spectrum, which also translates into a slightly better constraint on the surface pressure. In the case of perfect mass knowledge, the surface pressure has the tightest constraint. The surface pressure is a derived parameter and it is linked to the sum of the partial pressure of the single gases in the atmosphere. Therefore the planetary mass correlates with the atmosphere composition overall. For example, a looser constraint on the planetary mass results in a slightly higher variance on the posterior distribution of water vapor.

\begin{table*}[ht]
    \centering
    \caption{Atmospheric parameters used to simulate the Modern Earth-like (cloud free) scenario and the retrieval results for the five scenarios considered. The error bars of the retrieval median results correspond to the 68\% confidence interval (i.e., 1$\sigma$).}\label{tab:modern_cld}
    \begin{tabular}{lcccccc}\toprule
        Parameter & Input & Perfect $M_p$ & Unknown $M_p$ & 2D prior & Gaussian prior (30\%) & Gaussian prior (10\%) \\ \hline
        Log($P_0$) [Pa] (derived)   &  5.00 & 5.01$^{+0.02}_{-0.01}$ & 5.03$^{+0.08}_{-0.09}$ & 5.01$^{+0.09}_{-0.12}$ & 5.01$^{+0.07}_{-0.09}$ & 5.01$^{+0.04}_{-0.04}$ \\
        Log(VMR$_{H_2O}$)           & -2.52 & -2.56$^{+0.04}_{-0.04}$ & -2.57$^{+0.05}_{-0.06}$ & -2.56$^{+0.07}_{-0.07}$ & -2.56$^{+0.06}_{-0.06}$ & -2.56$^{+0.04}_{-0.04}$ \\
        Log(VMR$_{CH_4}$)           & -5.70 & -5.76$^{+0.55}_{-3.85}$ & -5.75$^{+0.54}_{-3.64}$ & -5.76$^{+0.54}_{-3.65}$ & -5.81$^{+0.60}_{-3.92}$ & -5.85$^{+0.63}_{-3.91}$ \\
        Log(VMR$_{CO_2}$)           & -3.40 & -5.54$^{+2.24}_{-4.41}$ & -5.38$^{+2.09}_{-4.48}$ & -5.71$^{+2.40}_{-4.30}$ & -5.73$^{+2.42}_{-4.33}$ & -5.54$^{+2.24}_{-4.50}$\\
        Log(VMR$_{O_2}$)            & -0.67 & -0.62$^{+0.08}_{-0.08}$ & -0.65$^{+0.10}_{-0.11}$ & -0.63$^{+0.11}_{-0.12}$ & -0.63$^{+0.10}_{-0.10}$ & -0.62$^{+0.09}_{-0.09}$ \\
        Log(VMR$_{O_3}$)            & -6.15 & -6.14$^{+0.02}_{-0.02}$ & -6.14$^{+0.02}_{-0.02}$ & -6.14$^{+0.02}_{-0.02}$ & -6.14$^{+0.02}_{-0.02}$ & -6.14$^{+0.02}_{-0.02}$ \\
        Log(VMR$_{N_2}$)            & -0.11 & -0.12$^{+0.02}_{-0.03}$ & -0.11$^{+0.03}_{-0.03}$ & -0.12$^{+0.03}_{-0.04}$ & -0.12$^{+0.03}_{-0.04}$ & -0.12$^{+0.02}_{-0.03}$\\
        $M_p [M_{\oplus}]$          &  1.00 & $-$ & 1.06$^{+0.22}_{-0.20}$ & 1.00$^{+0.25}_{-0.26}$ & 1.00$^{+0.19}_{-0.19}$ & 1.00$^{+0.08}_{-0.08}$ \\
        $R_p [R_{\oplus}]$          &  1.00 & 1.01$^{+0.01}_{-0.01}$ & 1.00$^{+0.01}_{-0.01}$ & 1.00$^{+0.01}_{-0.01}$ & 1.00$^{+0.01}_{-0.01}$ & 1.00$^{+0.01}_{-0.01}$ \\
        $\mu$[g/mol] (derived)      & 28.83 & 28.95$^{+0.21}_{-0.17}$ & 28.89$^{+0.23}_{-0.20}$ & 28.92$^{+0.27}_{-0.22}$ & 28.93$^{+0.24}_{-0.20}$ & 28.94$^{+0.21}_{-0.18}$ \\
        \hline
    \end{tabular}
\end{table*}

\subsection{Modern Earth-like}
Stepping up the complexity, we then took into consideration the Moden Earth case with the addition of a cloud deck.
The presence of a cloud deck introduces uncertainty in the retrieval process: clouds can obscure the Rayleigh scattering slope, which is the primary feature used to constrain the abundance of N\textsubscript{2}. Furthermore, clouds can obscure the planetary surface, complicating the determination of the planetary radius.

\begin{figure*}[ht]
    \centering
    \includegraphics[scale=0.6]{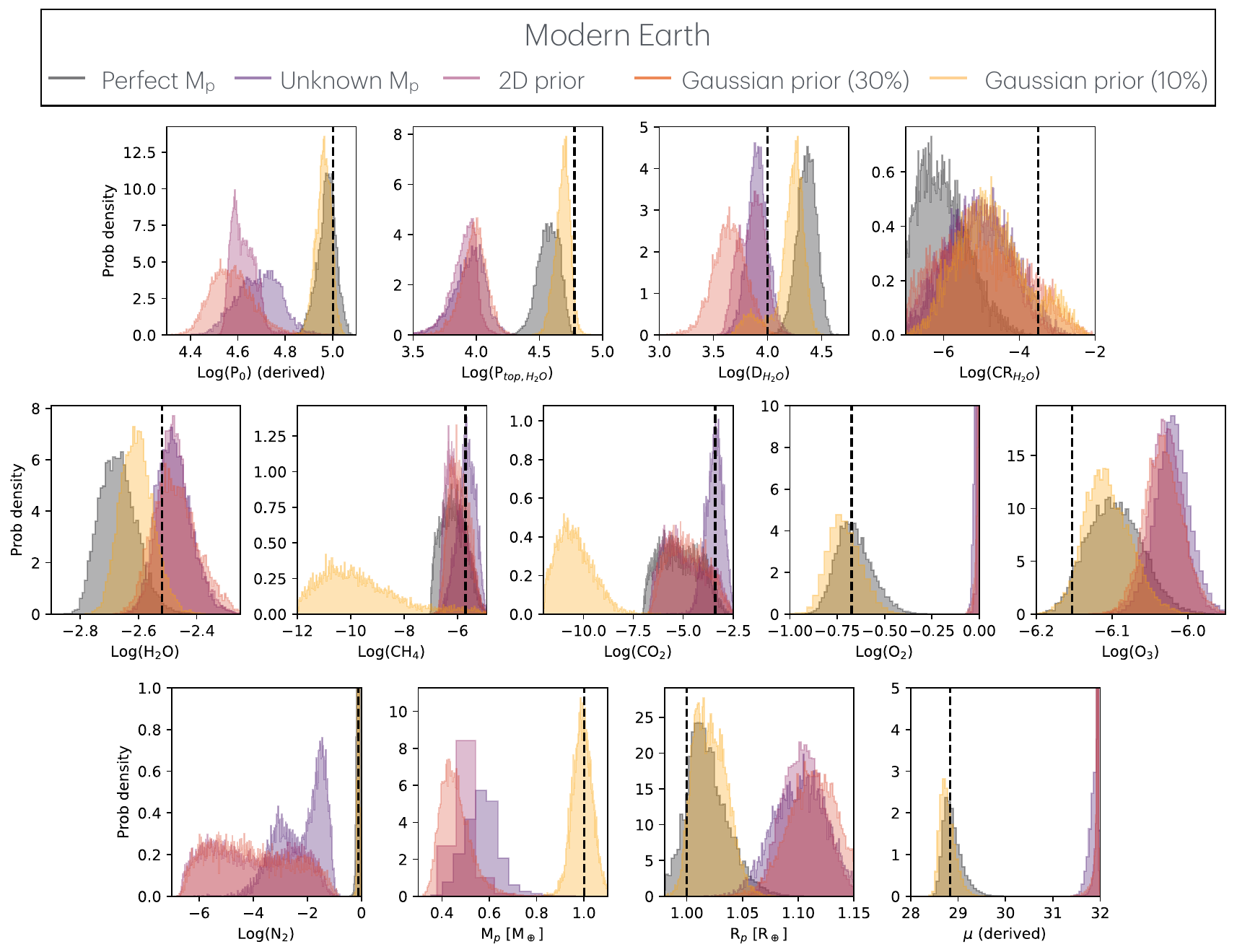}
    \caption{Posterior distribution functions of the Bayesian analysis on the Modern Earth-like reflected spectrum of all the scenarios defined in Section \ref{subsec:mass_2Dprior}.}
    \label{fig:1D_post_modern}
\end{figure*}

In the first scenario, we assume perfect knowledge of the planetary mass. This scenario serves as our baseline, against which subsequent scenarios are compared. As expected, all the true values have been accurately retrieved in the baseline scenario except for a small bias in the position of the cloud layer, but still within 3$\sigma$ (see Figure~\ref{fig:1D_post_modern}). The retrieved atmosphere is N\textsubscript{2}-dominated, with a significant amount of O\textsubscript{2}, followed by approximately 1\% H\textsubscript{2}O. The cloud deck is correctly retrieved within $3\sigma$ of the true value, and the planetary radius is well constrained at $1~R_{\oplus}$.

\begin{figure}[ht]
    \centering
    \includegraphics[scale=0.375]{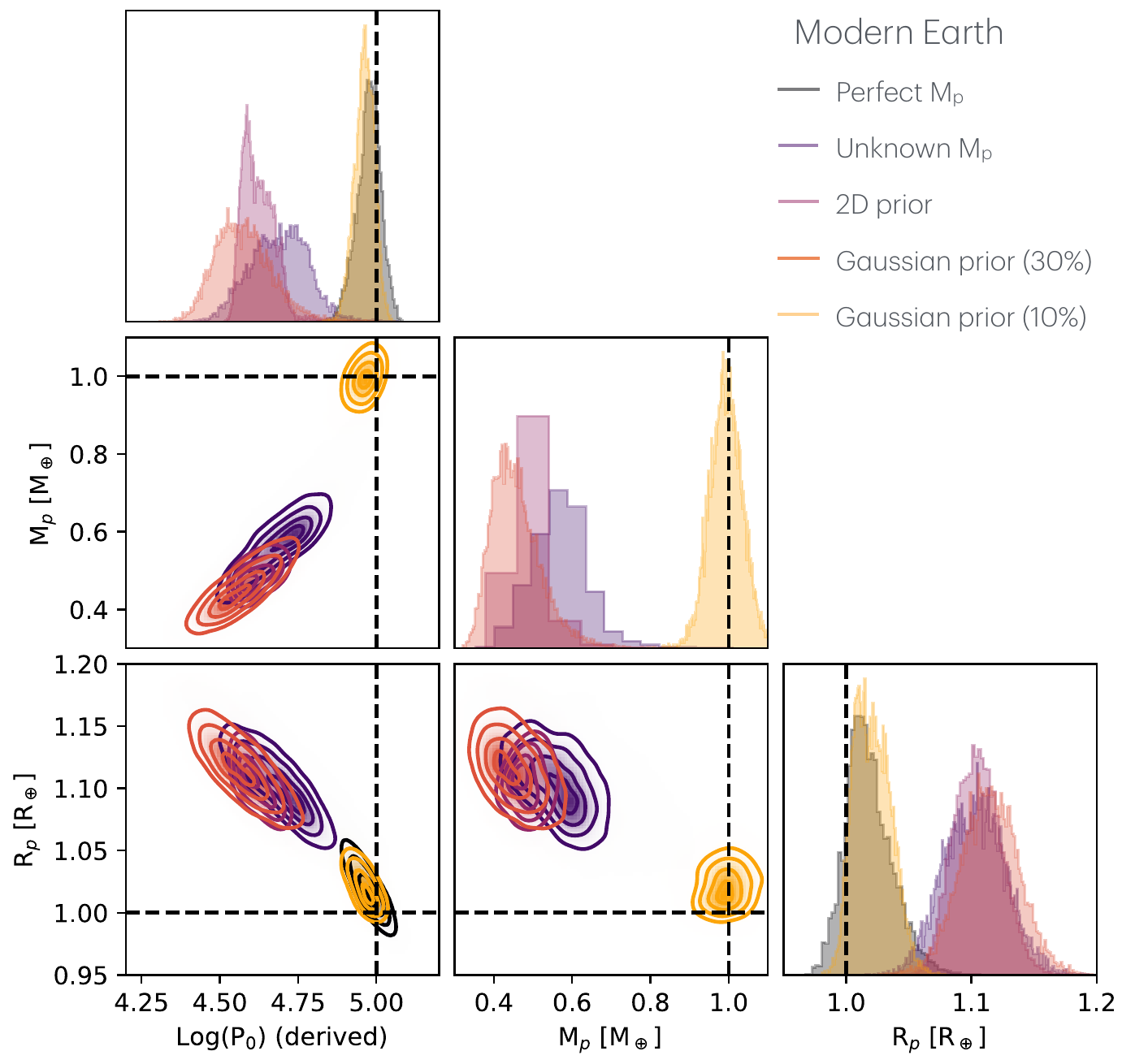}
    \caption{Posterior distribution functions for planetary mass, radius, and surface pressure for the Modern Earth-like scenario. The surface pressure is a proxy for the atmospheric gases collectively as the sum of their partial pressure is equal to the surface pressure.}
    \label{fig:mdrn_post}
\end{figure}

When the planetary mass is treated as a free parameter, we observe that the mass posteriors converge to values substantially smaller than the truth unless the prior mass constraints are $10\sigma$. 
In tandem with the biased mass estimate, the background gas is incorrectly identified as oxygen instead of nitrogen (Figure~\ref{fig:1D_post_modern}). This misidentification likely arises because the Rayleigh scattering slope at low wavelengths is similar for O$_2$ and N$_2$, and the slope is impacted by the planetary mass and can be obscured by clouds. While the other gases were mostly retrieved correctly (e.g., the true values bracketed by posterior distributions), we find a slight bias in the retrieved mixing ratio of ozone. Methane (CH\textsubscript{4}) and carbon dioxide (CO\textsubscript{2}) may, in some cases, be loosely constrained, as their abundances are low enough to not produce significant absorption features that can be fully interpreted by the retrieval. Additionally, the cloud deck was retrieved at a higher altitude by about an order of magnitude.

Again, we observe correlations between the planetary mass, radius, and surface pressure (Figure~\ref{fig:mdrn_post}). 
Here, the relationship appears linear when the surface pressure is expressed in logarithmic space and the mass in linear space. Similarly, we observe a comparable behavior between the planetary radius and surface pressure, which translates into a correlation between mass and radius (Figure~\ref{fig:mdrn_post}). Physically, this behavior can be explained by considering that a less massive planet has a lower surface gravity; thus, a lower pressure above the cloud deck is required to maintain the same atmospheric column in number density.

\begin{figure}[ht]
    \centering
    \includegraphics[scale=0.35]{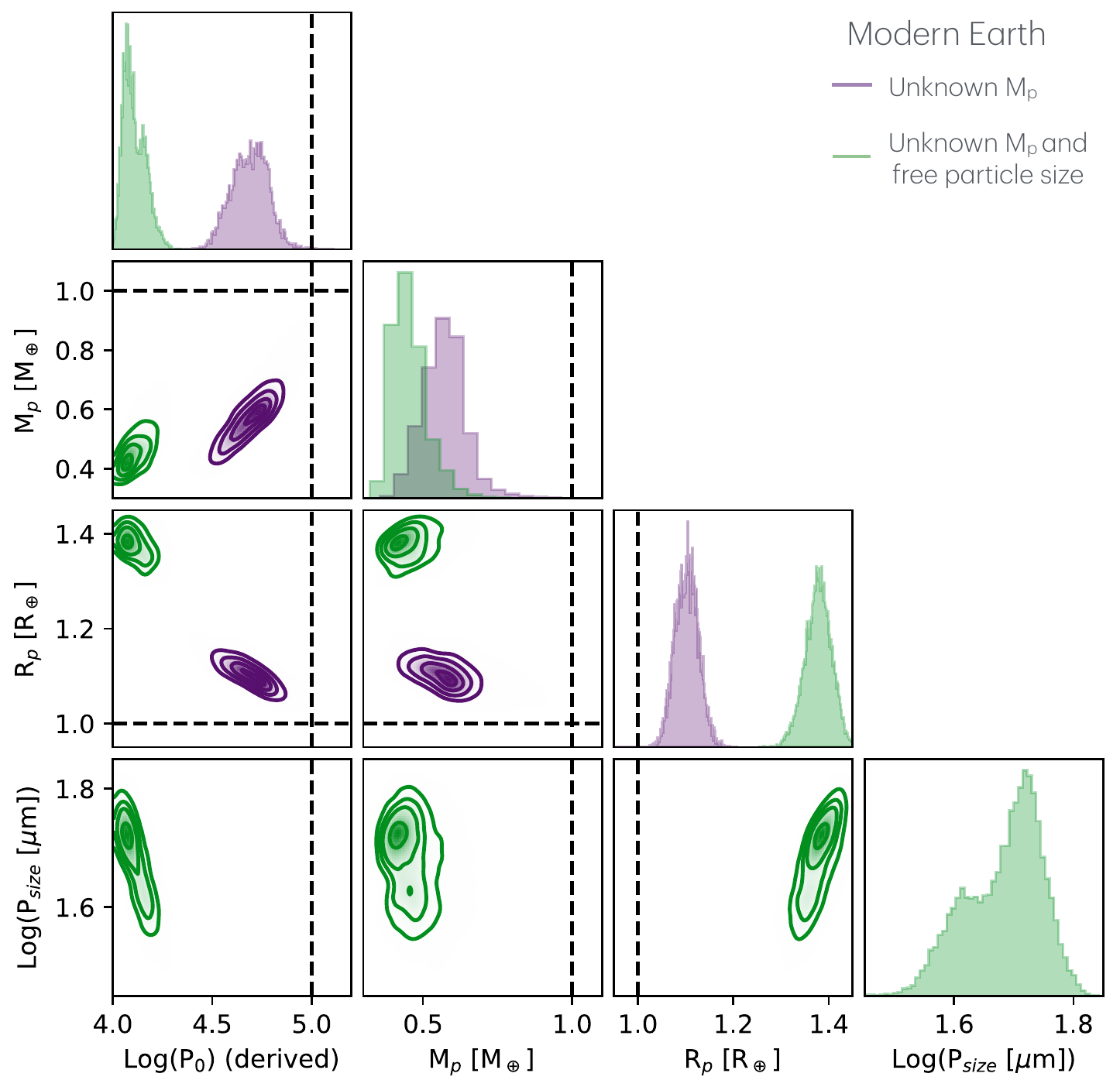}
    \caption{Posterior distribution functions for planetary mass, radius, surface pressure, and particle size for the Modern Earth-like scenario. The surface pressure is a proxy for the atmospheric gases collectively as the sum of their partial pressure is equal to the surface pressure.}
    \label{fig:mdrn_psize}
\end{figure}

Additionally, we also consider a variation of this scenario in which we also set the cloud particle size as a free parameter. 
When the particle size is retrieved to a constant average value throughout the atmosphere, it reveals additional effects in the constraints of mass, radius, and surface pressure (Figure~\ref{fig:mdrn_psize}). We observe that the posterior distribution of surface pressure and radius are now biased and the planetary mass is centered at slightly smaller values. The surface pressure and the radius are inversely correlated, i.e., the larger the planet, the lower the surface pressure.

A very similar behavior is observed when running the retrieval using other types of prior functions for the planetary mass, including a Gaussian prior with a 30\% standard deviation. Although conceptually the two-dimensional prior and the Gaussian prior are more stringent in terms of prior assumptions, the retrieved planetary mass remains consistent with that of the unknown mass prior scenario. Finally, when a Gaussian prior with a $10\%$ standard deviation is used for the planetary mass, the prior constraint becomes sufficient to correctly identify the atmospheric components and other free parameters (Figure~\ref{fig:1D_post_modern}). This result underlines how a tight prior knowledge of the mass is needed to correctly identify the background gas and the surface pressure on an Earth-like exoplanet.

\begin{table*}[ht]
    \centering
    \caption{Atmospheric parameters used to simulate the Modern Earth-like scenario and the retrieval results for the five scenarios considered. The error bars of the retrieval median results correspond to the 68\% confidence interval (i.e., 1$\sigma$).}\label{tab:modern}
    \begin{tabular}{lcccccc}\toprule
        Parameter & Input & Perfect $M_p$ & Unknown $M_p$ & 2D prior & Gaussian prior (30\%) & Gaussian prior (10\%) \\ \hline
        Log($P_0$) [Pa] (derived)   &  5.00 & 4.98$^{+0.03}_{-0.04}$ & 4.69$^{+0.09}_{-0.10}$ & 4.62$^{+0.06}_{-0.04}$ & 4.57$^{+0.09}_{-0.08}$ & 4.96$^{+0.03}_{-0.03}$\\
        Log($P_{top,H_2O}$) [Pa]    &  4.78 & 4.58$^{+0.08}_{-0.09}$ & 3.95$^{+0.11}_{-0.14}$ & 3.91$^{+0.08}_{-0.12}$ & 3.99$^{+0.09}_{-0.10}$ & 4.69$^{+0.05}_{-0.06}$\\
        Log($D_{cld,H_2O}$) [Pa]    &  4.00 & 4.36$^{+0.09}_{-0.09}$ & 3.91$^{+0.09}_{-0.09}$ & 3.86$^{+0.11}_{-0.14}$ & 3.65$^{+0.15}_{-0.15}$ & 4.24$^{+00.08}_{-0.14}$\\
        Log(CR$_{H_2O}$)            & -3.50 & -6.06$^{+0.72}_{-0.57}$ & -5.05$^{+0.89}_{-0.84}$ & -5.15$^{+0.81}_{-0.90}$ & -5.05$^{+1.30}_{-1.16}$ & -4.82$^{+1.0}_{-0.81}$\\
        Log(VMR$_{H_2O}$)           & -2.52 & -2.67$^{+0.07}_{-0.06}$ & -2.48$^{+0.06}_{-0.05}$ & -2.48$^{+0.06}_{-0.05}$ & -2.48$^{+0.08}_{-0.07}$ & -2.61$^{+0.06}_{-0.05}$\\
        Log(VMR$_{CH_4}$)           & -5.70 & -6.17$^{+0.51}_{-0.48}$ & -5.65$^{+0.33}_{-0.35}$ & -6.10$^{+0.48}_{-0.32}$ & -5.97$^{+0.42}_{-0.34}$ & -9.94$^{+1.67}_{-1.14}$\\
        Log(VMR$_{CO_2}$)           & -3.40 & -5.23$^{+1.22}_{-1.05}$ & -3.40$^{+0.41}_{-0.47}$ & -5.06$^{+1.42}_{-0.94}$ & -4.86$^{+1.11}_{-0.96}$ & -10.53$^{+1.01}_{-0.80}$\\
        Log(VMR$_{O_2}$)            & -0.67 & -0.67$^{+0.10}_{-0.08}$ & -0.01$^{+0.00}_{-0.01}$ & -0.001$^{+0.001}_{-0.001}$ & -0.001$^{+0.001}_{-0.001}$ & -0.72$^{+0.09}_{-0.08}$\\
        Log(VMR$_{O_3}$)            & -6.15 & -6.10$^{+0.04}_{-0.04}$ & -6.02$^{+0.02}_{-0.02}$ & -6.03$^{+0.02}_{-0.02}$ & -6.03$^{+0.03}_{-0.02}$ & -6.11$^{+0.03}_{-0.03}$\\
        Log(VMR$_{N_2}$)            & -0.11 & -0.11$^{+0.02}_{-0.03}$ & -2.01$^{+0.63}_{-1.17}$ & -4.13$^{+1.85}_{-1.62}$ & -4.25$^{+1.94}_{-1.44}$ & -0.09$^{+0.02}_{-0.03}$\\
        $M_p [M_{\oplus}]$          &  1.00 & $-$ & 0.57$^{+0.06}_{-0.07}$ & 0.49$^{+0.04}_{-0.03}$ & 0.45$^{+0.07}_{-0.05}$ & 0.99$^{+0.04}_{-0.04}$\\
        $R_p [R_{\oplus}]$          &  1.00 & 1.02$^{+0.02}_{-0.01}$ & 1.10$^{+0.02}_{-0.03}$ & 1.10$^{+0.02}_{-0.02}$ & 1.11$^{+0.02}_{-0.02}$ & 1.02$^{+0.02}_{-0.01}$\\
        $\mu$[g/mol] (derived)      & 28.83 & 28.85$^{+0.23}_{-0.15}$ & 31.92$^{+0.04}_{-0.13}$ & 31.95$^{+0.01}_{-0.02}$ & 31.95$^{+0.01}_{-0.02}$ & 28.75$^{+0.18}_{-0.13}$\\
        \hline
    \end{tabular}
\end{table*}

\subsection{Archean Earth-like}

Given the results of the Modern Earth-like scenario, we decided to test the same set of mass uncertainties on a different kind of atmosphere. The Archean Earth-like planet has considerably less oxygen and more CO$_2$ than the Modern Earth-like planet. This way we can test if the misinterpretation of the background gas is linked to O$_2$/N$_2$ correlations or if they are an effect of the mass uncertainty more generally. 

\begin{figure*}[ht]
    \centering
    \includegraphics[scale=0.6]{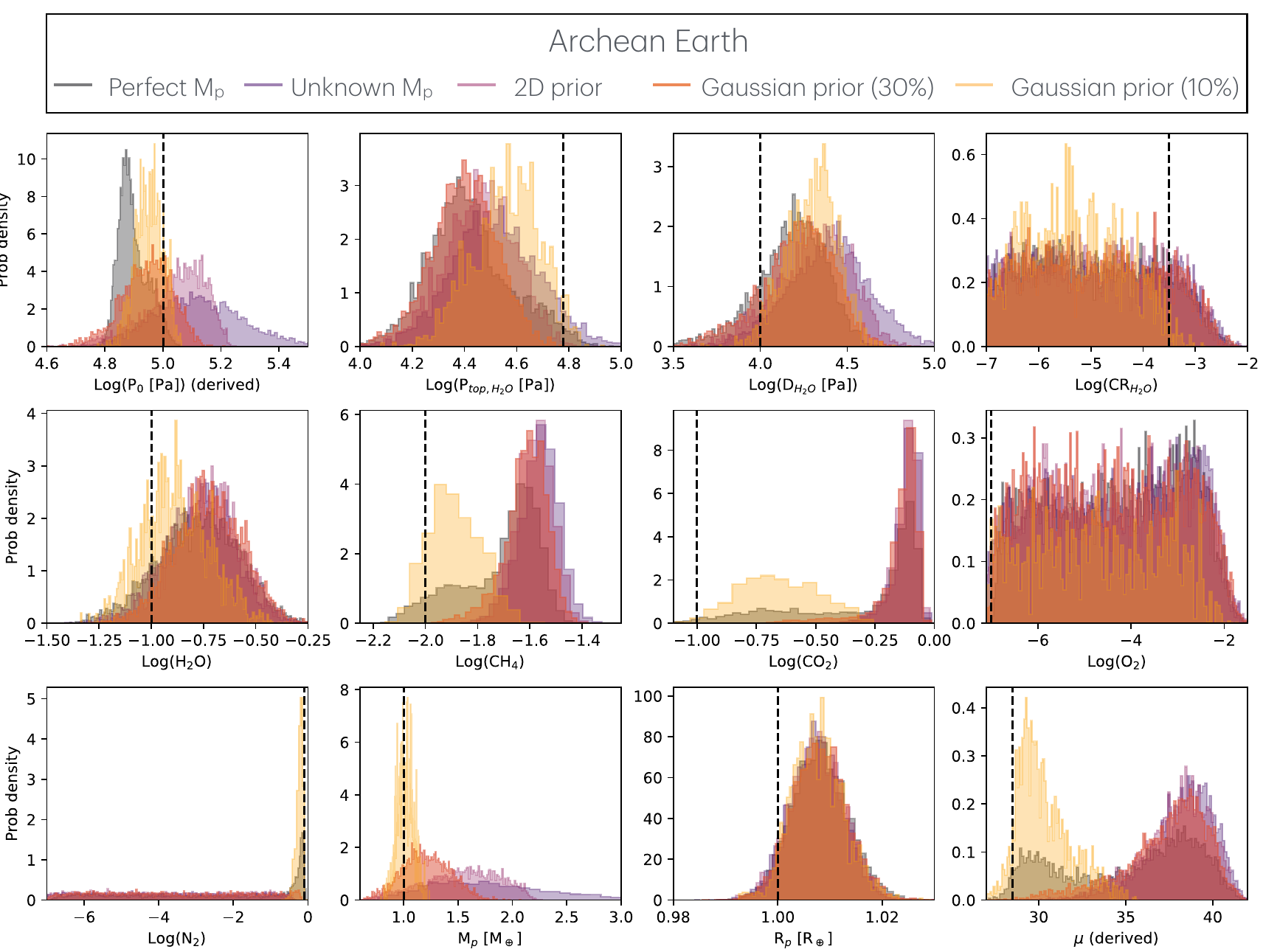}
    \caption{Posterior distribution functions of the Bayesian analysis on the Archean Earth-like reflected spectrum of all the scenarios defined in Section \ref{subsec:mass_2Dprior}.}
    \label{fig:1D_post_archean}
\end{figure*}

As in the case of the Modern Earth-like scenario, we started by fixing the planetary mass. In this case (Figure~\ref{fig:1D_post_archean}), we retrieved all the parameters correctly, i.e., the truth values are encompassed within the posterior distributions. Using this as the baseline scenario, we then included the mass as a free parameter and explored the different prior functions for the planetary mass described in Section \ref{subsec:method_setup}.

\begin{figure}[ht]
    \centering
    \includegraphics[scale=0.375]{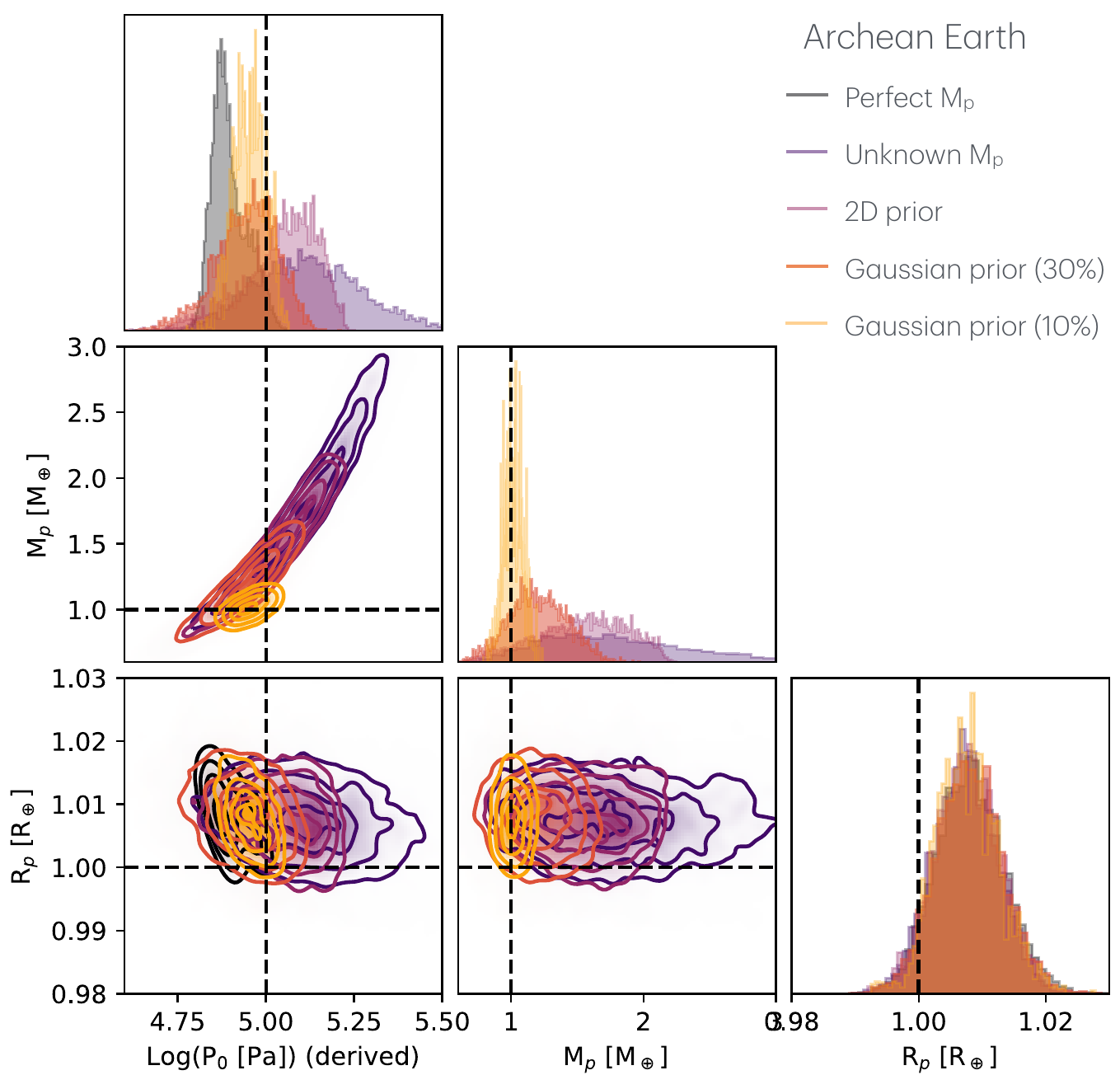}
    \caption{Posterior distribution functions for planetary mass, radius, and surface pressure for the Archean Earth-like scenario. Note that the surface pressure is a proxy for the atmospheric gases collectively as the sum of their partial pressure is equal to the surface pressure.}
    \label{fig:arch_post}
\end{figure}

Similarly to the Modern Earth scenario, we find that the background gas would be incorrectly identified if the planetary mass is not known to better than $\sim10\sigma$ (Figure~\ref{fig:1D_post_archean}). Here, the retrieval settled on CO$_2$ rather than N$_2$ as the background gas if the mass is not assumed to be well known. This shows that the misinterpretation of the background gas does not seems to be related to O$_2$ particularly but rather to how the retrieval interprets the gases that have strong spectral features versus the gases that do not. Misidentification of the dominant gas causes biases in the estimates of other gases. For example, CH$_4$ was retrieved slightly higher than the input value. H$_2$O was correctly constrained and O$_2$ was unconstrained as expected as there are no measurable spectral features with an input value of 10$^{-7}$. The cloud deck was overall correctly identified. The position is slightly higher and slightly deeper than the input cloud, but still within 3$\sigma$ from the input values.

Meanwhile, even though the mass was allowed to span up to 20M$_{\oplus}$, the retrieval converged to a small range around the true value, confirming that the spectrum indeed contains information about the planetary mass. With a flat prior on the mass, we find a broad posterior distribution centered at larger values (i.e., 1.77$_{-0.51}^{+0.80}$~M$_{\oplus}$), with the planetary radius retrieved correctly with a tight constraint around 1R$_{\oplus}$. Using the 2D prior on the planetary mass and radius tightens the mass constraints to 1.56$_{-0.40}^{+0.34}$~M$_{\oplus}$, as the mass now depends on the value of the radius. This result suggests that, in some cases, the spectrum alone can give planetary mass constraints to $\sim3\sigma$.

Giving a closer look to the correlations between key parameters (Figure~\ref{fig:arch_post}), we observe that the planetary mass and the surface pressure also in this case appear to be correlated. The correlation appears to be similar to the one observed in both previous atmospheric scenarios (see Figures~\ref{fig:mdrn_post_nocld} and \ref{fig:mdrn_post}), with a tighter mass constraint leading to a tighter constraint on the surface pressure. Unlike the Modern Earth-like with clouds scenario, we do not observe correlations between the planetary radius and the other two parameters.

\begin{figure}[ht]
    \centering
    \includegraphics[scale=0.35]{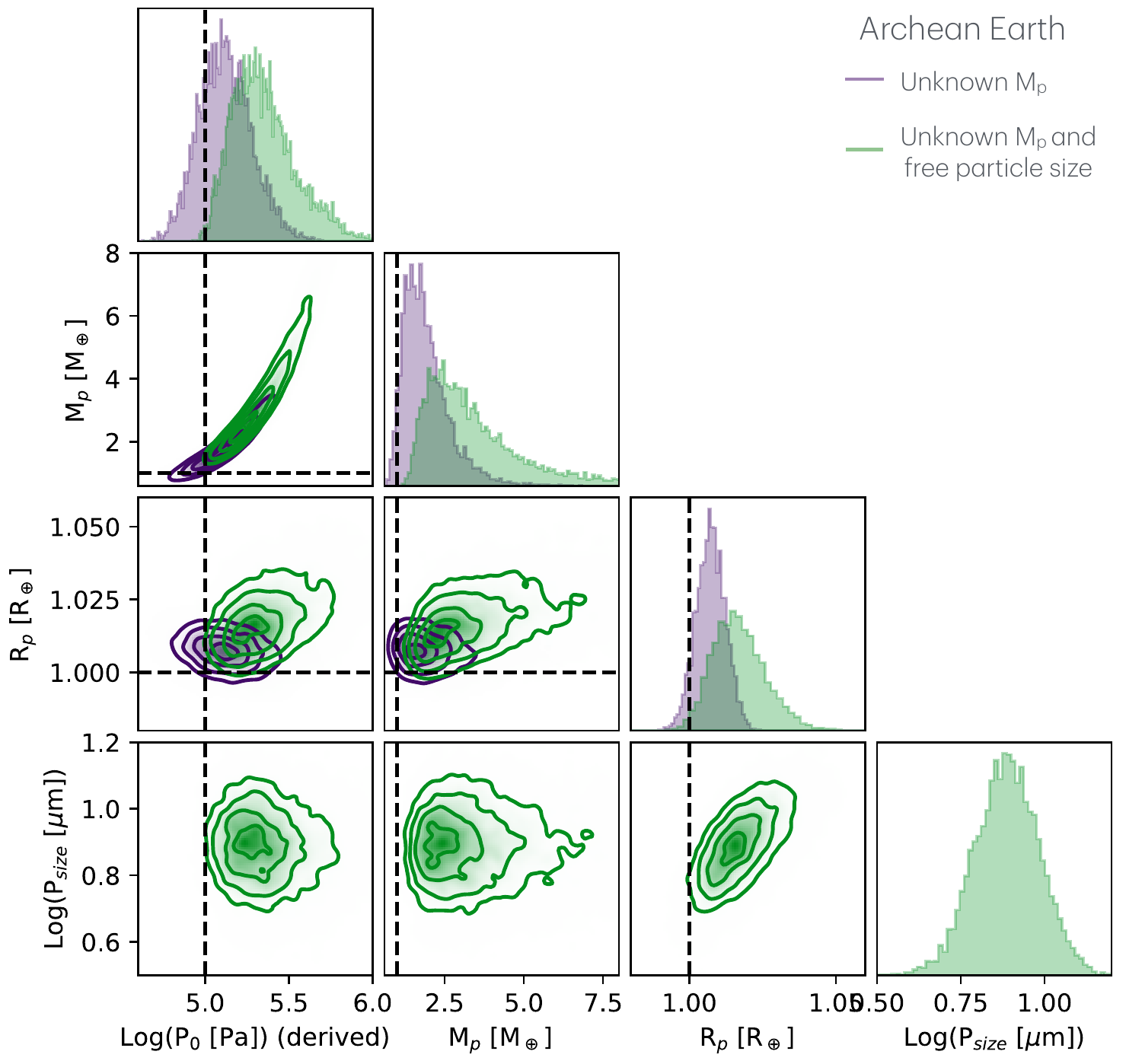}
    \caption{Posterior distribution functions for planetary mass, radius, surface pressure, and particle size for the Archean Earth-like scenario. The surface pressure is a proxy for the atmospheric gases collectively as the sum of their partial pressure is equal to the surface pressure.}
    \label{fig:arch_psize}
\end{figure}

Moreover, as in the case of the Modern Earth analog, we consider a variation of this first case for the Archean Earth analog by including the cloud particle size as a free parameter (Figure~\ref{fig:arch_psize}). Differently from the Modern Earth-like scenario, we do not observe biases, however we notice that the posterior distribution of surface pressure, mass, and radius have longer tails towards larger values, and the planetary mass can now stretch to $>6$ M$_{\oplus}$. While the retrieved value for the particle size largely agrees with the self-consistent formulation used to synthesize the spectrum in the first place, treating the cloud particle size as a free parameter notably degrades the mass constraints obtained from the spectrum.

\begin{table*}[ht]
    \centering
    \caption{Atmospheric parameters used to simulate the Archean Earth-like scenario and the retrieval results for the five scenarios considered. The error bars of the retrieval median results correspond to the 68\% confidence interval (i.e., 1$\sigma$).}\label{tab:archean}
    \begin{tabular}{lcccccc}\toprule
        Parameter & Input & Perfect $M_p$ & Unknown $M_p$ & 2D prior & Gaussian prior (30\%) & Gaussian prior (10\%)\\ \hline
        Log($P_0$) [Pa] (derived)   &  5.00 & 4.89$^{+0.07}_{-0.04}$ & 5.12$^{+0.16}_{-0.15}$ & 5.06$^{+0.09}_{-0.12}$ & 4.95$^{+0.08}_{-0.09}$ & 4.95$^{+0.05}_{-0.05}$\\
        Log($P_{top,H_2O}$) [Pa]    &  4.78 & 4.40$^{+0.18}_{-0.14}$ & 4.51$^{+0.17}_{-0.16}$ & 4.47$^{+0.13}_{-0.14}$ & 4.40$^{+0.12}_{-0.13}$ & 4.57$^{+0.13}_{-0.14}$\\
        Log($D_{cld,H_2O}$) [Pa]    &  4.00 & 4.19$^{+0.17}_{-0.22}$ & 4.40$^{+0.22}_{-0.24}$ & 4.32$^{+0.20}_{-0.23}$ & 4.23$^{+0.19}_{-0.23}$ & 4.31$^{+0.13}_{-0.17}$\\
        Log(CR$_{H_2O}$)            & -3.50 & -5.01$^{+1.42}_{-1.27}$ & -4.95$^{+1.52}_{-1.40}$ & -4.93$^{+1.45}_{-1.35}$ & -4.92$^{+1.39}_{-1.36}$ & -5.40$^{+1.09}_{-0.95}$\\
        Log(VMR$_{H_2O}$)           & -1.00 & -0.77$^{+0.19}_{-0.21}$ & -0.77$^{+0.15}_{-0.18}$ & -0.74$^{+0.15}_{-0.16}$ & -0.72$^{+0.16}_{-0.17}$ & -0.89$^{+0.17}_{-0.15}$\\
        Log(VMR$_{CH_4}$)           & -2.00 & -1.66$^{+0.09}_{-0.23}$ & -1.56$^{+0.07}_{-0.08}$ & -1.58$^{+0.07}_{-0.07}$ & -1.61$^{+0.07}_{-0.08}$ & -1.90$^{+0.11}_{-0.09}$\\
        Log(VMR$_{CO_2}$)           & -1.00 & -0.16$^{+0.07}_{-0.50}$ & -0.11$^{+0.03}_{-0.06}$ & -0.11$^{+0.03}_{-0.05}$ & -0.12$^{+0.04}_{-0.08}$ & -0.66$^{+0.19}_{-0.17}$\\
        Log(VMR$_{O_2}$)            & -7.00 & -4.23$^{+1.50}_{-1.80}$ & -4.31$^{+1.64}_{-1.85}$ & -4.39$^{+1.64}_{-1.71}$ & -4.42$^{+1.67}_{-1.68}$ & -6.68$^{+2.96}_{-3.19}$\\
        Log(VMR$_{N_2}$)            & -0.10 & -2.11$^{+1.92}_{-3.36}$ & -3.86$^{+2.34}_{-2.24}$ & -3.91$^{+2.34}_{-2.14}$ & -3.78$^{+2.56}_{-2.07}$ & -0.21$^{+0.07}_{-0.10}$\\
        $M_p [M{_\oplus}]$          &  1.00 & $-$ & 1.77$^{+0.80}_{-0.51}$ & 1.56$^{+0.34}_{-0.40}$ & 1.20$^{+0.24}_{-0.21}$ & 1.01$^{+0.08}_{-0.07}$\\
        $R_p [R_{\oplus}]$          &  1.00 & 1.01$^{+0.01}_{-0.01}$ & 1.01$^{+0.01}_{-0.01}$ & 1.01$^{+0.01}_{-0.01}$ & 1.01$^{+0.01}_{-0.01}$ & 1.01$^{+0.01}_{-0.01}$\\
        $\mu$[g/mol] (derived)      & 28.49 & 36.16$^{+2.92}_{-6.09}$ & 38.46$^{+1.55}_{-2.17}$ & 38.21$^{+1.71}_{-2.85}$ & 37.85$^{+1.59}_{-2.14}$ & 29.90$^{+1.93}_{-1.02}$\\
        \hline
    \end{tabular}
\end{table*}
\section{Discussion}\label{sec:discuss}
\subsection{Comparison with previous studies}

\cite{salvador2024mass} explored the impact of prior knowledge of planetary mass and orbital parameters on the atmospheric characterization of directly imaged Earth-like exoplanets in reflected light. The authors simulated observations of a modern Earth analog obtained by future missions like the Habitable Worlds Observatory. They performed atmospheric retrievals under varying levels of prior information, including scenarios with no prior constraints on planetary orbit and mass, known orbit, known orbit and mass, and only mass constrained.

The study finds that prior knowledge of orbit-related parameters, such as orbital distance and phase angle, significantly improves the determination of the planetary radius from reflected light observations, thereby helping to identify the planet's type and density. This improvement arises due to a strong degeneracy between the orbit and radius in the planet-to-star flux ratio. However, additional prior knowledge of the planet's mass does not notably enhance radius determination or atmospheric characterization in that study. 



In this study, we assume that the planetary orbit has been constrained within reasonable errobars so that the position of the planet in its orbit would be known. Once the orbit is known, the characterization observation can be scheduled and the reflected spectrum of the planet can be observed. In this work, we focused more on the interaction between the planetary mass and the atmospheric composition. 
The results shown in this paper agree with \cite{salvador2024mass} in a sense that the spectrum can generally constrain the planetary radius when the orbital configuration is known. We do see that the spectrum contains some information about the mass, and when imposing the 2D prior introduced in this paper, the mass may even be constrained from the spectrum alone in some cases (Figure~\ref{fig:arch_post}). However, as a crucial revelation, our results suggest that the ability to retrieve the background gas hinges on highly precise prior mass constraints. 

\subsection{Identification of the background gas}
Misidentifying the background gas can be a substantial issue as it can render a habitable planet (moderate concentration of O$_2$) into a post-runaway, inhospitable world (massive O$_2$-dominated atmosphere). Indeed, the identification of O$_2$ as a biosignature relies on the knowledge that the background atmosphere is not O$_2$ dominated \citep{wordsworth2014abiotic,harman2015abiotic,meadows2017o2bio,spross2021n2o2}.

In this work, we showed that the lack of precise planetary mass prior knowledge will likely prevent the background gas from being correctly identified, even with a spectrum that covers from the UV to the NIR. In the case of the Modern Earth analog with clouds, N$_2$ is misidentified as O$_2$, and in the case of the Archean Earth-like planet, the N$_2$ is misidentified as CO$_2$. In both cases, the common factor is an active absorbing gas versus a gas that does not show absorption features in the considered wavelength range. Not having a mass constraint equates to not having a surface gravity constraint, and this affects the calculation of the atmospheric scale height. The retrieval algorithm seeks to balance the difference in the scale height with different mean molecular masses. In the case of Modern Earth, the two most abundant gases have close mean molecular mass (28 for N$_2$ and 32 for O$_2$), and the mass constraints must be stringent enough to distinguish between the two. For the Archaen Earth analog, instead, the difference between N$_2$ and CO$_2$ in terms of mean molecular mass is greater (28 for the former and 44 for the latter), and we see that different constraints on the planetary mass are indeed reflected in the posterior distribution of N$_2$ and CO$_2$. A strong fit of the Rayleigh scattering will help in the identification of the background gas (see Sec.~\ref{sec:mdrn_earth_nocld}). However, the presence of clouds tends to mute the Rayleigh feature, making it more challenging to uniquely associate it with the background gas.


\subsection{Towards 10\% mass uncertainty}

Radial Velocity (RV) observations are the most successful tool to date for determining the masses of small exoplanets. Measurements of a star's radial velocity (RV) made using stable, high resolution, optical or near infrared spectrographs can be used to determine the orbital parameters of the exoplanets it hosts \citep{Lovis2010}. Time series measurements of Doppler shifts in the star’s spectral lines that are caused by the gravitational influence of an orbiting planet allow for the derivation of the planet’s minimum mass (M$_p$ $sin(i)$), where $i$ is the inclination of the planet’s orbital plane relative to the line of sight. 

Most RV discovered planets to date, where no precise period and orbital phase priors are provided by accompanying transit detections, have RV semi-amplitudes K \textgreater 1 m s$^{-1}$. This limit is set primarily by the presence of stellar phenomena such as star spots, plage, and granulation which can deform stellar absorption features and obscure Keplerian Doppler signals \citep[see, e.g.,][]{Meunier2010,CollierCameron2019}. The RV semi-amplitudes of an Earth mass planet orbiting at the Earth Equivalent Irradiation Distance around likely HWO target stars \citep{Mamajek2024} ranges from 5 – 50 cm s$^{-1}$, notably smaller than the current state of the art. Achieving the 10\% mass uncertainty recommended in this study will therefore require that the uncertainties of the relative RV measurements be $\leq$ 10 cm s$^{-1}$ over time scales of years to decades so that systematic errors do not dominate the planet signals \citep{Luhn2023}.

While 10-sigma mass measurements ($\sigma_{m_{\rm{pl}}}$ / m$_{\rm{pl}} \geq$ 10)  of Earth analogs have not yet been demonstrated, the development of a new generation of Extreme Precision RV (EPRV) instruments including ESPRESSO \citep{Pepe2021}, NEID \citep{Schwab2016} , EXPRES \citep{Petersburg2020} and KPF \citep{Gibson2024} has improved the precision of single RV measurements to $\sim$30 cm s$^{-1}$. And dedicated treatment of instrument systematics and stellar activity in long baseline RV time series from the previous generation of stabilized RV instruments has demonstrated Doppler Sensitivity at the $50 - 60$ cm s$^{-1}$ level over timescales of $1-2$ years \citep[see, e.g.,][]{Cretignier2023, John2023}. 

Further improvements to both instrument performance and stellar variability modeling / mitigation will be required in the coming years to advance extreme precision RV (EPRV) capabilities, but these recent results show promising forward momentum. We note, however, that the radial velocity method alone measures only the projected mass of the planet, leaving its true mass uncertain. We also note that stars above the Kraft break (Teff $\geq$ 6250K) will not be amenable to EPRV measurements because their fast rotation periods produce significant rotational broadening thereby decreases the amount of Doppler information content contained within their spectra \citep{Beatty2015}.

Astrometry is another powerful technique for determining the masses of exoplanets. By precisely tracking the star's motion on the plane of the sky over time, astrometry could provide a direct measurement of the planet's mass without the $\sin i$ ambiguity inherent in radial velocity methods, as it is sensitive to the inclination of the orbit \citep{unwin2008astrometry}. In the context of the HWO, astrometry could play a crucial role in directly measuring the masses of small terrestrial planets \citep{malbet2012neat, quirrenbach2014carmenes}. The combination of astrometric measurements from HWO with its direct imaging capabilities can yield both the mass and orbital parameters of Earth-sized planets in the habitable zones of nearby stars, enhancing our understanding of their potential habitability \citep{astro2020national}.

\section{Conclusion}\label{sec:conclusion}
In this study, we explored the impact of planetary mass uncertainties on the atmospheric retrieval of terrestrial exoplanets observed in the reflected light. Utilizing an enhanced version of \exorelr, we incorporated several upgrades, including cloud fraction modeling, composition-dependent Rayleigh scattering, an adaptive vertical grid, partial pressure sampling, and the inclusion of planetary mass as a free parameter with various prior probability functions.
We also introduced a method to more realistically calculate the errorbars of the simulated spectrum. This new routine takes into consideration the star photon noise, exozodi approximation, and spectral resolution of the simulated data; and analytically calculates the signal to noise per spectral element.

Our retrieval analyses were conducted on three atmospheric scenarios: a cloud-free modern Earth-like atmosphere, a modern Earth-like atmosphere with a cloud deck, and an Archean Earth-like atmosphere. We considered different prior knowledge of the planetary mass, ranging from perfectly known mass to scenarios with varying degrees of uncertainty.

The results indicate that precise knowledge of the planetary mass is needed for accurate atmospheric characterization of small rocky planets. When the planetary mass is known within 10\% uncertainty, the retrievals successfully identify the background gas and constrain atmospheric parameters, even in the presence of clouds. However, with less constrained or unknown mass, we observed significant biases in the retrievals, particularly in the misidentification of the dominant atmospheric gas. For instance, nitrogen was incorrectly replaced by oxygen or carbon dioxide, which could lead to incorrect assessments of planetary habitability and biosignatures.

These biases arise because uncertainties in planetary mass affect the determination of surface gravity and atmospheric scale height, leading the retrieval algorithm to compensate by adjusting the atmospheric composition. While some information about the planetary mass can be inferred from the reflected spectrum alone, it is insufficient to accurately constrain the mass and the bulk atmospheric composition without additional prior information.

In this work, we did not explore the impact of mass uncertainty on characterizing larger planets (e.g., super-Earths versus sub-Neptunes), nor higher spectral resolution or signal-to-noise ratios. Nonetheless, our simulations highlight the importance of achieving precise mass measurements of small terrestrial exoplanets—ideally within 10\% uncertainty—through methods such as Extreme Precision Radial Velocity or astrometry, especially in the context of missions like the Habitable Worlds Observatory. Accurate mass constraints are important to reliably characterize exoplanet atmospheres and to correctly interpret potential biosignatures.
Future observational strategies should prioritize obtaining accurate mass measurements to improve the characterization of exoplanetary atmospheres and to better assess their potential habitability.

\section*{Acknowledgments}
    We thank Vanessa P. Bailey and Bertrand Mennesson for helpful discussion
    The High Performance Computing resources used in this investigation were provided by funding from the JPL Information and Technology Solutions Directorate.
    This research was carried out at the Jet Propulsion Laboratory, California Institute of Technology, under a contract with the National Aeronautics and Space Administration (80NM0018D0004).

\section*{Software}
    \noindent \textsc{Numpy} \citep{oliphant2015numpy}, \textsc{Scipy} \citep{virtanen2020scipy}, \textsc{Astropy} \citep{astropy2013,astropy2018,astropy2022}, \textsc{Matplotlib} \citep{hunter2007matplotlib}, \textsc{MultiNest} \citep{feroz2009multinest,buchner2014multinest}, \textsc{mpi4py} \citep{dalcin2021mpi4py}.






\appendix
\section{Noise Model}\label{sec:noise}
There are multiple sources of noise in a coronagraphic image or spectrum including detector noise; host star residual speckles after point spread function (PSF) subtraction; and shot noise from the planet, local zodi, exozodi, and the host star raw speckles. The relative strengths of these various noise sources will depend on the the instrument architecture and the observational target. For example, coronagraphs require active wavefront control and therefore achieve higher performance on brighter host stars; and planet radius, orbital distance, albedo, and host star type all contribute to the observed flux ratio.

The noise model used for this study represents a simplistic limiting case with very good starlight suppression and low detector noise, in which photon noise from the astrophysical scene dominates. We follow the analytic prescription outlined in \citet{robinson2016characterizing}. We can assume exozodi dominate over local zodi, and neglect local zodi. Dark current is also ignored, assuming that a future exoplanet imaging mission will have good enough detectors that the dark current will be lower than photon noise at all wavelengths, as is already the case for the Roman CGI in broadband imaging \citep{morrissey2023photon}. The speckle residuals are likewise neglected. This leaves the shot noise from the planet and the exozodi as the only remaining noise sources.

Future mission concept studies would use more realistic noise models that are representative of the chosen architecture. However, as we normalize the noise to the desired SNR, the main difference between this model and the "full" model is the color of noise, rather than the magnitude. Running the retrieval with different colors of noise was outside the scope of this research.

The planet's photoelectron count rate at the detector within an aperture is:
\begin{equation} \label{eq:planet}
    c_{p,\lambda} = F_{p,\lambda} \frac{\lambda}{hc} \pi \left(\frac{D}{2}\right)^2 \Delta\lambda \mathcal{T}_\lambda f_{ap} \propto F_{p,\lambda} \lambda \Delta\lambda,
\end{equation}
where $\frac{\lambda}{hc}$ is the photon energy, $\Delta\lambda$ is the bandwidth of the filter or spectral resolution element, $D$ is the telescope diameter, and $\mathcal{T}_\lambda$ is the effective system throughput (detector quantum efficiency and optical throughput), and $f_{ap}$ is the fraction of planet light within the aperture (sometimes also called the "core throughput"; in the coronagraphic case this also accounts for losses due to the coronagraph masks and wavefront control). We assume $\mathcal{T}_\lambda$ and $f_{ap}$ are wavelength-independent.

For extended surface brightness sources, such as exozodi, the count rate is
\begin{equation}
    c_{ez,\lambda} = F_{ez,\lambda} \frac{\lambda}{hc} \pi \left(\frac{D}{2}\right)^2 \Delta\lambda \mathcal{T}_\lambda \Omega \propto F_{s,\lambda} \lambda^3 \Delta\lambda,
\end{equation}
where $\Omega$ is the aperture size. In this case we assume an aperture proportional to the PSF width, therefore $\Omega \propto \lambda^2$. We assume exozodi are gray ($F_{ez,\lambda} \propto F_{s,\lambda} $). 

To calculate the total noise, we combine the noise sources with appropriate weights. The total count rate for photon noise is given by
\begin{equation}
    c_{r,\lambda} = c_{p,\lambda} + \alpha c_{ez,\lambda}
\end{equation}

The SNR for a given exposure time can then be calculated as
\begin{equation}
    SNR = \frac{c_{p,\lambda} t_{exp}}{\sqrt{ (c_{p,\lambda} + \alpha c_{ez,\lambda})t_{exp}}}
\end{equation}
Substituting the definitions for each term allows us to investigate the proportionality with wavelength and spectral bin width:
\begin{equation}
    SNR = \beta F_{p,\lambda} \sqrt{\frac{\lambda \Delta \lambda}{F_{p,\lambda} + \alpha F_{s,\lambda} \lambda^2}}
\end{equation}
$\alpha$ and $\beta$ are weights chosen to scale the noise terms and the overall SNR to the desired values. According to \citet{hu2021noise}, the flux of the exozodi should be roughly half that of the planet at 0.7$\mu m$, assuming an Earth-like planet in the HZ of an FGK star at 6pc, and assuming that the system has 3 zodi (as is the assumption adopted by various mission concept studies and yield analyses \citep{gaudi2020habex, seager2019starshade, stark2019exoearth}). Thus $\alpha$ is chosen such that $F_{p,\lambda}=2\alpha F_{s,\lambda} \lambda^2$ at 0.75$\mu m$. $\beta$ is chosen to scale the noise such that an SNR of 20 is achieved at 0.75$\mu m$. This wavelength was chosen as it is on the continuum for all three scenarios analyzed in this paper, and it is next to the $O_2$ absorption feature at 0.76$\mu m$.

\begin{figure}[hbt]
    \centering
    \includegraphics[width=0.48\textwidth]{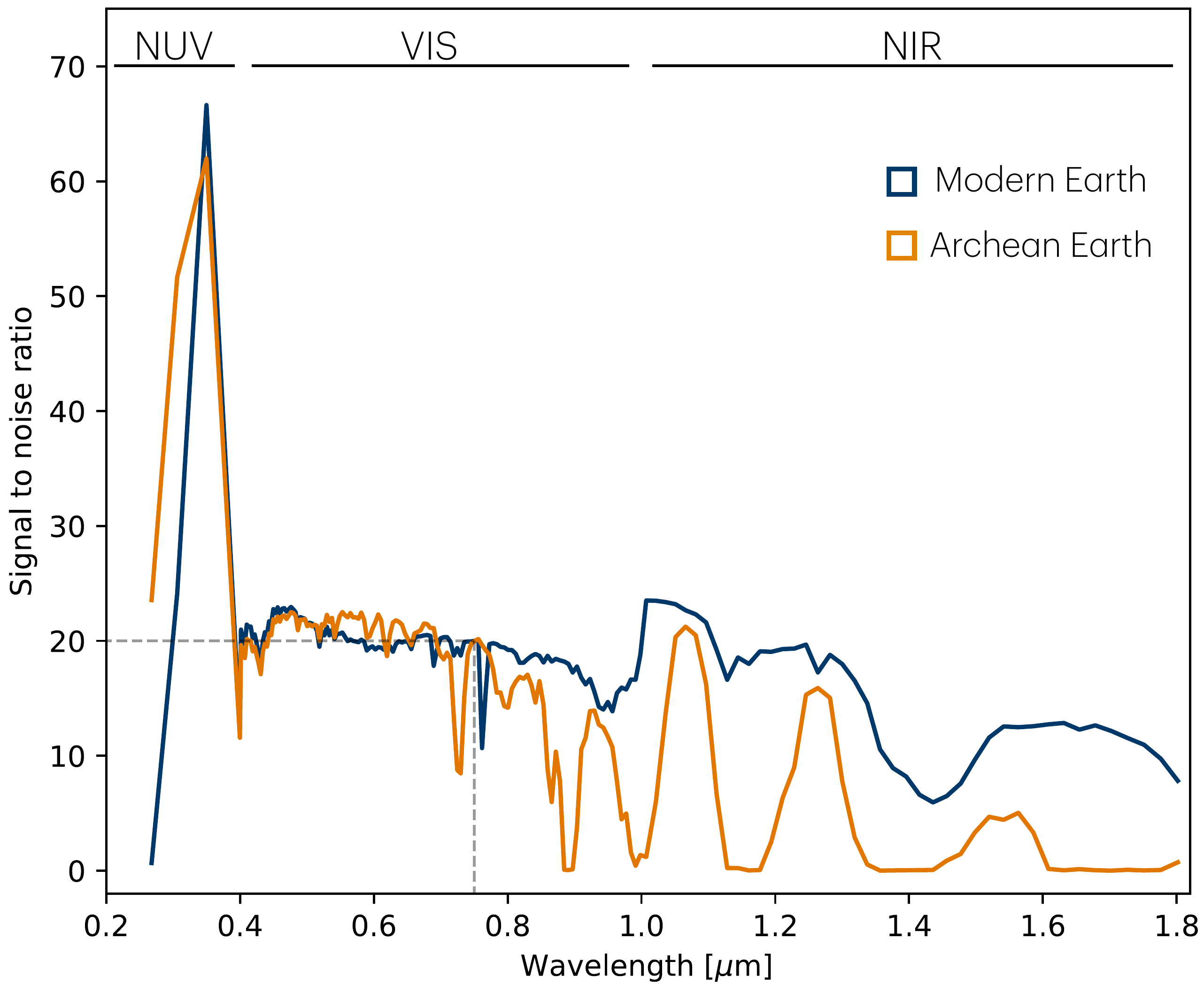}
    \caption{SNR vs. $\lambda$ for the modern and Archean Earth scenarios. The SNR is set to be 20 at 0.75$\mu m$. The jump at 1$\mu m$ is due to the fact that in the NIR the spectral resolution decreases from R=140 to R=70. Similarly, the jump at 0.4$\mu$m is due to the drop in spectral resolution in the NUV, i.e., R=7.}
    \label{fig:SNR}
\end{figure}

The resulting SNR vs. wavelength for the modern and Archean Earth scenarios are shown in \autoref{fig:SNR}.

\bibliography{references, bib}{}
\bibliographystyle{aasjournal}



\end{document}